# Quantum Compressive Sensing: Mathematical Machinery, Quantum Algorithms, and Quantum Circuitry

**Kyle M. Sherbert [1], Naveed Naimipour [1,2], Haleh Safavi [1], Harry C. Shaw [1,*] and Mojtaba Soltanalian [2]**

[1] NASA Goddard Space Flight Center, Greenbelt, ML 20771, USA; kmsherbert@yahoo.com (K.M.S.); naveed.naimipour@nasa.gov (N.N.); haleh.safavi@nasa.gov (H.S.)
[2] Department of Electrical and Computer Engineering, University of Illinois at Chicago, Chicago, IL 60607, USA; msol@uic.edu
* Correspondence: harry.c.shaw@nasa.gov

**Abstract:** Compressive sensing is a sensing protocol that facilitates the reconstruction of large signals from relatively few measurements by exploiting known structures of signals of interest, typically manifested as signal sparsity. Compressive sensing's vast repertoire of applications in areas such as communications and image reconstruction stems from the traditional approach of utilizing non-linear optimization to exploit the sparsity assumption by selecting the lowest-weight (i.e., maximum sparsity) signal consistent with all acquired measurements. Recent efforts in the literature consider instead a data-driven approach, training tensor networks to learn the structure of signals of interest. The trained tensor network is updated to "project" its state onto one consistent with the measurements taken, and is then sampled site by site to "guess" the original signal. In this paper, we take advantage of this computing protocol by formulating an alternative "quantum" protocol, in which the state of the tensor network is a quantum state over a set of entangled qubits. Accordingly, we present the associated algorithms and quantum circuits required to implement the training, projection, and sampling steps on a quantum computer. We supplement our theoretical results by simulating the proposed circuits with a small, qualitative model of LIDAR imaging of earth forests. Our results indicate that a quantum, data-driven approach to compressive sensing may have significant promise as quantum technology continues to make new leaps.

## 1. Introduction

Recent advances in quantum computing technology have motivated new, "quantum" perspectives in machine learning. As a prime example, the classical idea of a "Boltzmann machine", which represents the probability distribution of a training set, can be adapted to that of a "Born machine", which gives complex-valued amplitudes whose square moduli are that same probability distribution [1]. The Born machine is easily understood as a quantum state that can be realized using qubits on a quantum computer, and correlations in the training set can be realized using entanglement between qubits [1,2]. Robust and large-scale quantum computing is, however, still a distant technological reality. Tensor-network-based machine learning methods have ascended to a special niche, because they are classically-tractable mathematical structures which nevertheless approximate quantum states (e.g., Born machines) in a well-defined and practical way [3]. The Matrix Product State (MPS) tensor network is a particularly attractive model, as its linear structure accommodates an especially efficient implementation using the Density Matrix Renormalization Group algebra [4]. MPS has a long history in the numerical study of 1D quantum systems, and it is now being applied to the field of machine learning: Stoudenmire and Schwab [5] proposed a supervised classification algorithm using MPS. Han et al. [6] extend upon their methods to implement an unsupervised generative modeling algorithm. More recently, Ran et al. [7] proposed a communication protocol, christened Tensor Network Compressed Sensing (TNCS), in which Alice sends Bob only a partial message, and Bob recovers the full message with the help of an MPS tensor network. All of these approaches use the language and formalism of quantum Born machines, while restricting implementation to an approximation of a fully quantum state describable with an MPS. The purpose of this work is to provide algorithms for implementing and extending the methodology of these past approaches on bona-fide quantum computers.

The "Compressed Sensing" in TNCS is the name of a signal processing protocol, also known as *compressive sensing* or *compressive sampling*. A sensing protocol is one in which an $n$-dimensional signal is sampled by $m$ different measurements. In general, the measurement procedure can be described with an $m \times n$ "sensing matrix" $A$, such that the measurement vector $x \in R^m$ and signal vector $y \in R^n$ are related by the equation $x = Ay$. The problem of reproducing the original signal from the set of measurements appears to be equivalent to finding the left inverse $A^{-1}$, so that $y = A^{-1}x$; this requires $m \geq n$ at a *minimum*, and even more measurements may be required to ensure $A$ is non-singular. However, in *compressive* sensing, one takes $m \ll n$. In the classical setting, this is possible only under the assumption that the signal $y$ is *sparse*, so that it can be reconstructed via a minimization over the weight (i.e., the $l_0$ norm) of all possible signals, constrained so as to be consistent with the measurements taken [8–11]. Sparsity of a signal is basis-dependent: mathematically, one may always factor the sensing matrix $A = \Phi\Psi$ such that the transformed signal $y' = \Psi y$ is sparse. Even without *knowing* the appropriate transformation $\Psi$, compressive sensing can still be completed successfully as long as $\Phi$ possesses a certain mathematical property called "restricted isometry." "Restricted isometry" essentially requires that the vector $\Phi y$ has approximately the same norm as the original signal $y$, for all sparse signals $y$ [12,13]. Candès and Wakin [11] emphasize that randomly generated matrices do satisfy this requirement with overwhelming probability, both before and after factorization. In fact, one common way to approach the compressive sensing reconstruction problem is through convex optimization. This particular approach allows for said problem to be solved by a *linear programming* solver and can account for measurements corrupted by noise [14–18].

Duarte et al. [19] propose an engineering solution to implement random measurement for compressive sensing, using a single-pixel camera in conjunction with a micro-array of mirrors which can be randomly configured to redirect light into or away from the camera. Each measurement is the total intensity of light corresponding to a different configuration of the mirrors; it is thus equivalent to the integration over a random selection of pixels in a typical CCD image. This mode of compressive sensing has been realized in LIDAR geoimaging [20], which will be further discussed in our work. In addition to LIDAR geoimaging, compressive sensing has garnered an impressive reputation in the practical world with applications including underwater imaging [21] and microwave imaging [22], as well as communication networks such as wireless sensor networks [23,24], handling of big data for IoT [25], information security [26,27], and much more.

The *quantum* approach to compressive sensing adopts a somewhat different strategy. In TNCS [7], Alice and Bob share a tensor network representing a Born machine trained on the set of $n$-bit messages Alice is inclined to send. Alice identifies her message, which is analogous to the signal $y$ in compressive sensing. She then selects a subset of $m$ bits to send to Bob, either randomly or targeting those bits which yield the most information in the tensor network. This step is most analogous to the measurement step in compressive sensing. Finally, Bob reconstructs the full message with the help of the tensor network. We will describe this part of the algorithm in more detail in Section 2.

We find the relation of TNCS to compressive sensing very attractive, with the potential to be expanded upon. In particular, the TNCS measurement step has two major differences from traditional compressive sensing. First, Alice performs the measurement with full knowledge of the original signal; this makes it more akin to a compression algorithm rather than a *sensing* algorithm. Second, the measurement selects $m$ bits of an $n$-bit message, restricting the sensing matrix to one that is constructed only from standard basis vectors.

In this paper, we adapt TNCS from a quantum-inspired tensor network communication protocol to a fully-quantum sensing protocol, considering especially the context of LIDAR geoimaging. Let Mother Nature herself take the role of Alice, transmitting to our detector whatever sculpture of life and earth she is inclined to create. We offer a novel, gradient-free training procedure which can train a Born machine on, for example, known patterns in geological structure, or on previously obtained LIDAR images. We also demonstrate how arbitrary sensing matrices, such as those consistent with a single-pixel camera, may be translated into the TNCS formalism. Moreover, while both Han et al. and Ran et al. suggest the *possibility* of a quantum implementation of their algorithms, we propose several specific quantum protocols to implement our algorithm on a set of entangled qubits, permitting a much more powerful representation of highly correlated data.

## 2. Quantum Compressive Sensing

Our Quantum Compressive Sensing protocol is summarized in the following steps:

1. Training
2. Measurement
3. Projection
4. Sampling

In the *training* step, an $n$-qubit Born machine $|\Psi\rangle$ is prepared, representing the distribution of possible signals $y$. In the context of TNCS, $y$ may be the message Alice wishes to send to Bob. In the context of LIDAR geoimaging, each element of $y$ may represent the amount of foliage at a particular coordinate in space, so that we might interpret $y$ as an image with $n$ pixels. The Born machine maps each possible signal $y$ to a complex amplitude $\langle y|\Psi\rangle$ whose square modulus is the probability that $y$ is observed. We discuss the pixel–qubit mapping in more detail in Section 3, and training procedures in Section 4.

In the *measurement* step, $m$ distinct measurements are made on an unknown $n$-dimensional signal $y$, as described by a classical $m \times n$ sensing matrix. The measurement outcomes obtained are $x = Ay$. In the context of TNCS, $x$ may be the partial message which Alice sends to Bob. In the context of LIDAR geoimaging, each element of $x$ may be the total intensity reflected into a single-pixel camera by a randomly-assigned configuration of micro-mirrors. Both $x$ and $A$ are considered as inputs to the next step of our protocol.

In the *projection* step, the Born machine $|\Psi\rangle$ is projected onto a state $|\Psi_x\rangle$ which spans only those states consistent with the measurement outcome $x$. This step involves applying a non-unitary operator onto a quantum state, which is highly non-trivial. The majority of our paper addresses this problem with several different strategies in Section 5.

In the *sampling* step, all qubits are measured, obtaining a binary string representing a probabilistic estimate $y'$ of the original signal $y$. If the original signal $y$ is non-binary, the reconstruction $y'$ is at best an approximation, but it is (ideally) qualitatively sufficient to convey the required information (e.g., the geological structure being imaged, or the thesis of Alice's message). Optionally, the sampling process may be repeated many times to generate a sequence of binary signals $y$ whose statistics could be reverse-mapped onto a non-binary image. However, because both the projection and sampling steps irrevocably alter the state of the machine, $|\Psi\rangle$ would have to be re-trained after each sample. Therefore, we focus mainly on "single-copy" sampling throughout this work, although we acknowledge that multiple trials are generally required in both the training and projection steps.

## 3. Pixel–Qubit Mapping

Before diving into the details of our protocol, we must first describe our methodology for pixel–qubit mapping, in which we map a signal onto a quantum sate. Let the signal $y$ be an $n$-dimensional vector $y \in Y \subset R^n$, where $Y$ is the set of all signals we may expect to measure. We would like to prepare a Born machine $|\Psi\rangle$ on the set $Y$, implemented as a quantum state of $n$ qubits. We must therefore have a procedure for mapping a signal $y \in Y$ onto a quantum state $|y\rangle$, an element of the $d^n$-dimensional Hilbert space characterizing $n$ qubits. In this paper, we restrict ourselves to $d = 2$ (the ubiquitous "qubit") and to $n' = n$, meaning we require one qubit for each "pixel" in our image. This choice accommodates an entanglement-free *product state* representation of $|y\rangle$, where the state of each qubit depends only on the corresponding pixel $y_i$. The intuitive choice for binary signals is to map onto the computational basis state: if $y_i = 0$, the corresponding qubit has the state $|0\rangle$ and if $y_i = 1$, it has the state $|1\rangle$. A real-valued pixel may be mapped onto a linear combination of the two basis states. Stoudenmire and Schwab [5] use the following choice as a prototypical example:

$$|y_i\rangle \to \cos\left(\frac{\pi}{2} y_i\right)|0\rangle + \sin\left(\frac{\pi}{2} y_i\right)|1\rangle. \tag{1}$$

The use of sin and cos functions is convenient, because they automatically ensure a normalized qubit state. Further, assuming $y$ is normalized such that each pixel $y_i \in [0, 1]$, this mapping is sensible in that the closer a pixel is to 0 or 1, the higher the amplitude that the corresponding qubit is at for the corresponding state, and a pixel value $y_i = 0.5$ corresponds to an equal superposition of states.

We also consider a generalization to Equation (1) which designates an arbitrary pixel value $p$ as the "midpoint" between 0 and 1, rather than 0.5:

$$|y_i\rangle \to \cos\left[\frac{\pi}{2} f_p(y_i)\right]|0\rangle + \sin\left[\frac{\pi}{2} f_p(y_i)\right]|1\rangle, \tag{2}$$

where the function $f_p$ is given by

$$f_p(x) = \frac{1}{2}\left[1 + \frac{2}{\pi}\arctan\left[\tan\left(\pi\left(x - \frac{1}{2}\right)\right) - \tan\left(\pi\left(p - \frac{1}{2}\right)\right)\right]\right]. \tag{3}$$

The function $f_p : [0, 1] \to [0, 1]$ satisfies $f_p(0) = 0$, $f_p(p) = 0.5$, and $f_p(1) = 1$. It reduces to $f_p(x) = x$ for $p = 0.5$, and for $p \neq 0.5$ it is equivalent to a smooth non-linear rescaling of $y_i$ before applying the mapping in Equation (2). Figure 1 provides a visualization of $f_p(x)$ when different values of $p$ are applied. The state specified by Equation (2) for an arbitrary $n$-dimensional signal $y$ is a product state efficiently prepared by $n$ parallel single-qubit gates, as shown in Figure 2.

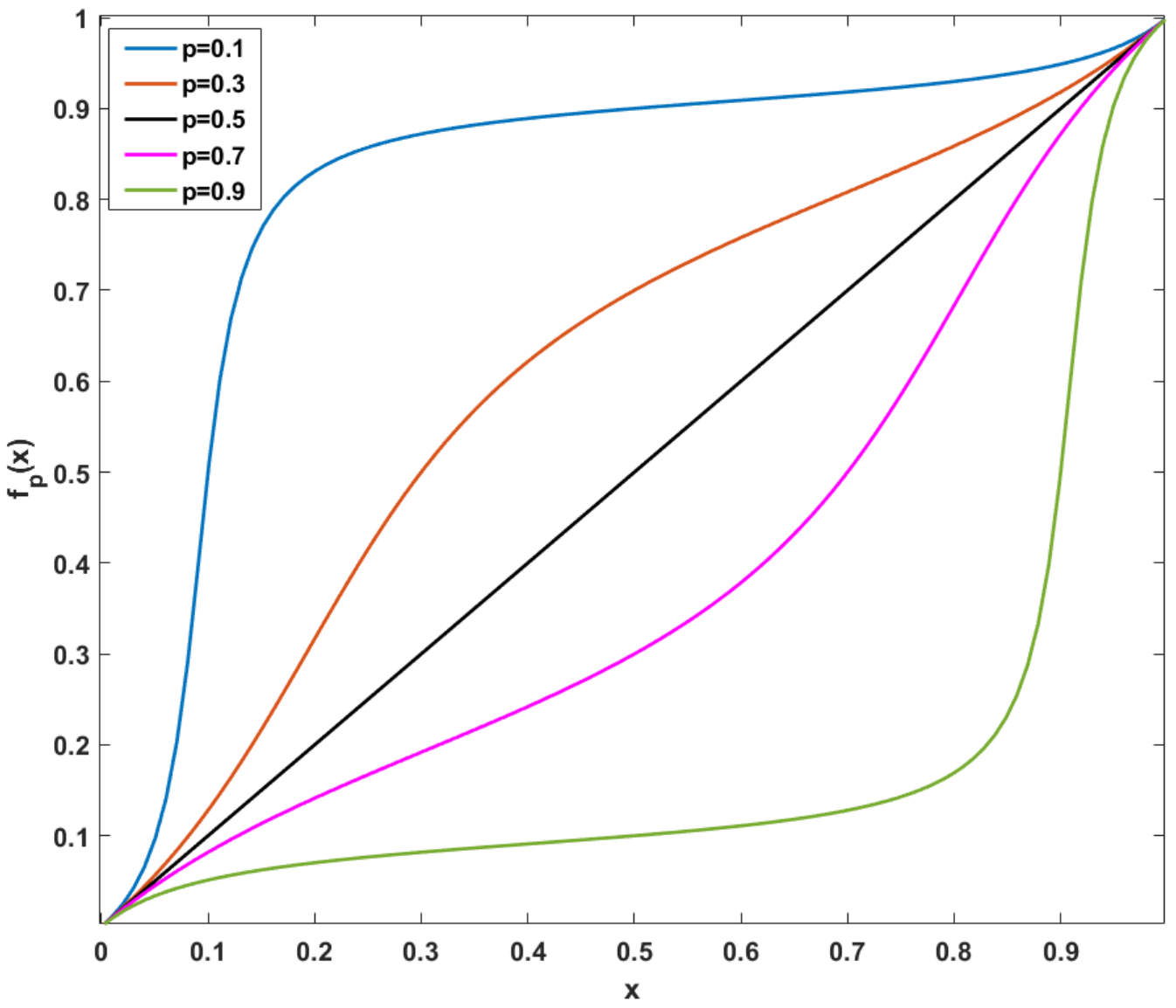

**Figure 1.** Visualization of Equation (3) as a function of $x$ when the "midpoint" $p = 0.1, 0.3, 0.5, 0.7, 0.9$.

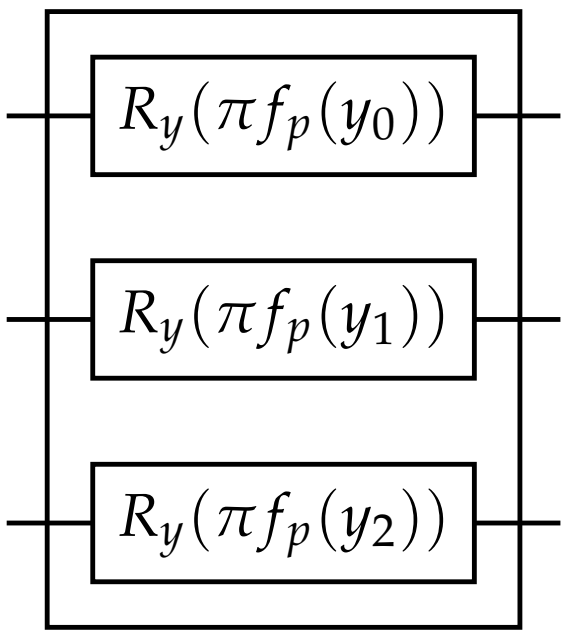

**Figure 2.** A simple quantum circuit to map an $n$-dimensional vector $y$ onto $n$ qubits. The quantum gate $R_y(\theta)$ is defined as $\exp\left(-i\frac{\theta}{2}\hat{Y}\right)$, where $\hat{Y}$ is the Pauli Y spin operator.

A clear disadvantage of both mappings proposed in Equations (1) and (2) is that they do not take full advantage of the Hilbert space of a single qubit. One could encode additional information into a complex phase difference between the $|0\rangle$ and $|1\rangle$ amplitudes. One area of further research is to develop alternative mappings accommodating higher entanglement or a more compact use of qubits.

## 4. Training

We now consider quantum algorithms to prepare $n$ qubits into a Born machine $|\Psi\rangle$ describing the set $Y$. Practically speaking, the Born machine is constructed in the machine learning sense—that is, on a representative *training set* $T \subset Y$, while we assume that $|\Psi\rangle$ generalizes usefully onto the entire population $Y$. A typical approach is to prepare $|\Psi\rangle$ variationally, applying a unitary *variational form* $V(\boldsymbol{\theta})$ to the initial state $|\mathbf{0}\rangle$ to prepare an *ansatz* $|\Psi(\boldsymbol{\theta})\rangle = V(\boldsymbol{\theta})|\mathbf{0}\rangle$. One then makes measurements on the ansatz to determine the expectation values of some cost function $C(\boldsymbol{\theta})$. The parameters $\boldsymbol{\theta}$ are tuned through a classical algorithm such as gradient descent until the cost function $C$ is minimized. The cost function is selected such that its minimum corresponds to a Born Machine which well represents the training set $T$. For example, the Negative Log Likelihood (NLL) function used in previous works is given as [6,7]:

$$C(\boldsymbol{\theta}) = -\frac{1}{|T|}\sum_{y\in T} \ln|\langle y|\Psi(\boldsymbol{\theta})\rangle|^2. \tag{4}$$

The variational form $V(\boldsymbol{\theta})$ should in principle be capable of preparing any ansatz $|\Psi\rangle$ in the Hilbert space of $n$ qubits. In gate-model quantum computing, any unitary operation must be decomposed into single- and two-qubit operations, and a fully general variational form may require an exponential number of operations. Therefore, one may choose to use a randomized circuit which merely approximates the Hilbert space. Layered circuits alternating between single-qubit rotations and entanglement between adjacent qubits have been shown to sample uniformly from the Hilbert space provided sufficient layers, making them promising candidates. However, recent theoretical efforts show that quantum variational algorithms are extremely vulnerable to the phenomenon of *barren plateaus*, in which large tracts of the parameter space are essentially flat, resulting in optimizations which either fail to converge or which converge at entirely arbitrary points [28,29].

These difficulties motivate us to explore an alternative training procedure, which does not require gradient descent or any such optimization process at all. Let us model the Born machine $|\Psi\rangle$ to be a *superposition* of each element in $Y$, in some sense a "quantum average" (Ran et al. [7] use this term in a similar way):

$$|\Psi\rangle = \frac{1}{\sqrt{|Y|}}\sum_{z=0}^{|Y|-1}|y_z\rangle. \tag{5}$$

If all $y \in Y$ are orthogonal, the square amplitude $|\langle y|\Psi\rangle|^2 = 1/|Y|$ is precisely the probability of retrieving $y$ from $Y$, as desired. It should be noted we assume all $y$ are uniformly distributed. In the more general case, we consider $Y$ a collection which may contain multiple copies of the same $y$, so that it is summed over more than once. This will not hold for a set $Y$ whose elements are not orthogonal, but this is advantageous: overlap between qualitatively similar images permits us to construct $|\Psi\rangle$ only from the training set $T$, and still achieve meaningful amplitudes for $y \notin T$.

There is no simple quantum operation which can deterministically prepare the sum of two arbitrary non-orthogonal quantum states. To see this, note that adding the states $|0\rangle$, $-|0\rangle$ (the same but for opposite phases) must result in the unnormalizable state 0, which cannot be prepared by a unitary operation. There is, however, a procedure by which we may yet achieve our goal. The quantum circuit presented in Figure 3 is a diagrammatic representation of a series of quantum operations acting on the computational basis state |000...>. The individual quantum operations, or *gates*, can be found in any standard quantum information textbook (e.g., [30]); the overall action of the entire quantum circuit prepares the following state:

$$|\Psi\rangle = \frac{1}{\sqrt{|T|}} \sum_{z=0}^{|T|-1} |z\rangle \sum_{z'=0}^{|T|-1} (-1)^{z \cdot z'} |y_{z'}\rangle, \tag{6}$$

where $|z\rangle$ represents the computational basis state of a "control register" with $\log_2 |T|$ qubits, $|y_z\rangle \equiv U_z|\mathbf{0}\rangle$, and the notation $z \cdot z'$ represents the inner product of the *binary expansions* of integers $z$ and $z'$. By measuring all qubits in the control register, we can project the state register into a single linear combination. If we happen to measure all control qubits in the basis state 0, corresponding to $z = 0$, we have prepared the desired linear combination modeling Equation (5).

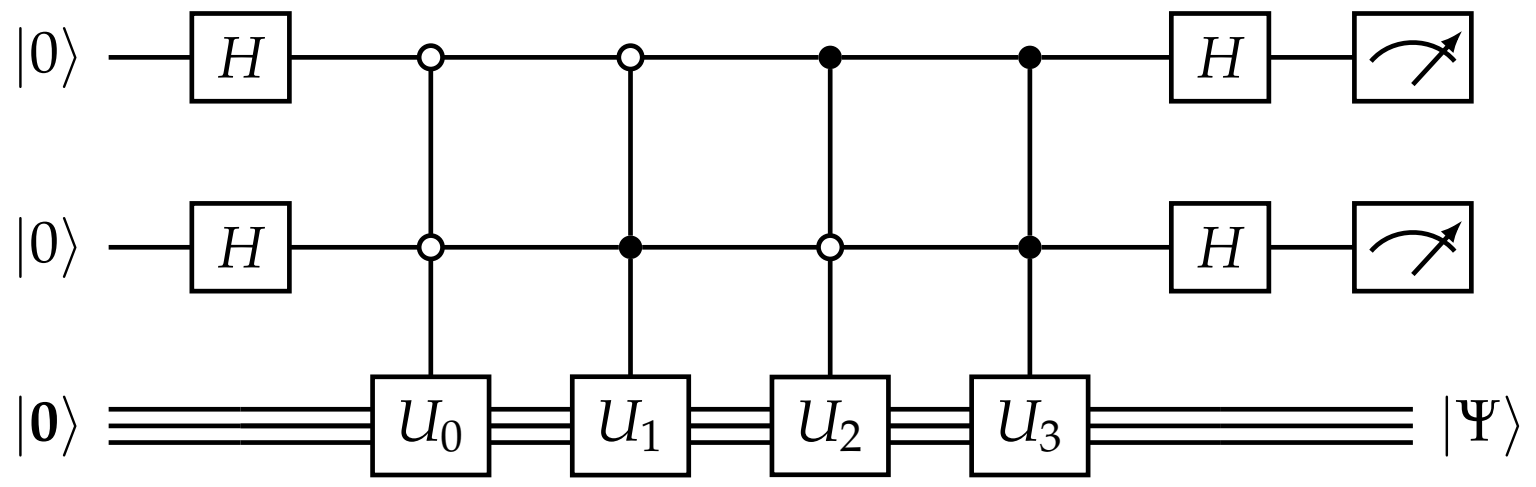

**Figure 3.** A quantum circuit to prepare the "quantum average" of the signals $y_i \equiv U_i|\mathbf{0}\rangle$. The circuits $U_i$ are the state-preparation circuits given in Figure 2. The quantum gate $H$ is the Hadamard gate. Dots on a wire represent controlled operations: a closed dot has a control value of $|1\rangle$ while an open dot has a control value of $|0\rangle$. Meters at the end of a circuit represent measurement.

The probability $P(z = 0)$ of successfully measuring all control qubits in the basis state 0 is given by $\langle 0|\rho_z|0\rangle$, where $\rho_z$ is the "reduced density matrix" of the control register obtained by tracing out all degrees of freedom in the state register. We may write it as $\rho_z = \mathrm{Tr}_y|\Psi\rangle\langle\Psi|$, where $\mathrm{Tr}_y$ is a partial trace, defined in any standard quantum information textbook. Substituting its definition into $P(z = 0)$, we find the probability to prepare our desired linear combination is $P(z = 0) = \sum_y (\langle 0|\langle y|)|\Psi\rangle\langle\Psi|(|0\rangle|y\rangle)$. Substituting in Equation (5), the probability simplifies to:

$$P(z = 0) = \frac{1}{|T|} + \frac{2}{|T|^2} \sum_{i<j} \mathrm{Re}[\langle y_i|y_j\rangle]. \tag{7}$$

In contrast with our previous discussions, this demonstrates a clear *advantage* of the qubit mappings described in Equations (1) and (2): under these mappings, the second term in this sum is strictly non-negative, meaning $P(z = 0) \geq 1/|T|$.

Therefore, the circuit in Figure 3 must be executed only $O(|T|)$ times in order to successfully prepare the quantum-average state. Since the circuit depth is also $O(|T|)$, the total run-time of this training procedure is $O(|T|^2)$ and requires no gradient descent or numerical optimization.

## 5. Projection

We now consider how to project our Born machine $|\Psi\rangle$ onto the state $|\Psi_x\rangle$ spanning only those signals $y$ consistent with a given measurement outcome $x = Ay$, where $A$ is some given $m \times n$ sensing matrix. Mathematically, projection is achieved via an idempotent ($\Pi^2 = \Pi$) but non-unitary projection operator $\Pi$. Non-unitary transformations are highly non-trivial to implement on a quantum computer, and this step, in addition to the sheer number of qubits, poses one of the main obstacles to implementing TNCS in a quantum computer. We *may* achieve a projection by measuring $m$ of the $n$ qubits, collapsing each one to a basis state 0 or 1. Entanglement between qubits ensures that the un-measured qubits have also collapsed to a corresponding state. However, in order to prepare the *desired* state $|\Psi_x\rangle$, we must repeat the training and measurement steps until the measured qubits' values are consistent with the classically observed measurement outcome $x$.

Consider the following straightforward example, in which the sensing matrix $A$ measures three individual pixels of the six-dimensional signal $y$.

$$A = \begin{bmatrix} 0 & 0 & 1 & 0 & 0 & 0 \\ 1 & 0 & 0 & 0 & 0 & 0 \\ 0 & 0 & 0 & 0 & 1 & 0 \end{bmatrix} \tag{8}$$

Let us assume that the result of our classical measurement yields $x = \begin{bmatrix} 1 & 0 & 1 \end{bmatrix}$, meaning the original signal $y$ is consistent with the vector $y = \begin{bmatrix} 0 & ? & 1 & ? & 1 & ? \end{bmatrix}$. Now we measure the first, third, and fifth qubits of the trained Born machine $|\Psi\rangle$. In order to implement the required projection, we must measure 0, 1, and 1 respectively. If we measure any other combination of 0's and 1's on these three qubits, we must discard this $|\Psi\rangle$, train another, and try again. However, once we have successfully measured 0, 1, and 1, we may then measure the remaining three qubits to fill in the question marks in $y$.

In the common case where the sensing matrix $A$ does *not* measure individual pixels, we must generalize our procedure. In fact, we offer four distinct projection protocols, each with its own technical challenges. As a reference, a visual overview of the protocol as a whole can be found in Figure 4.

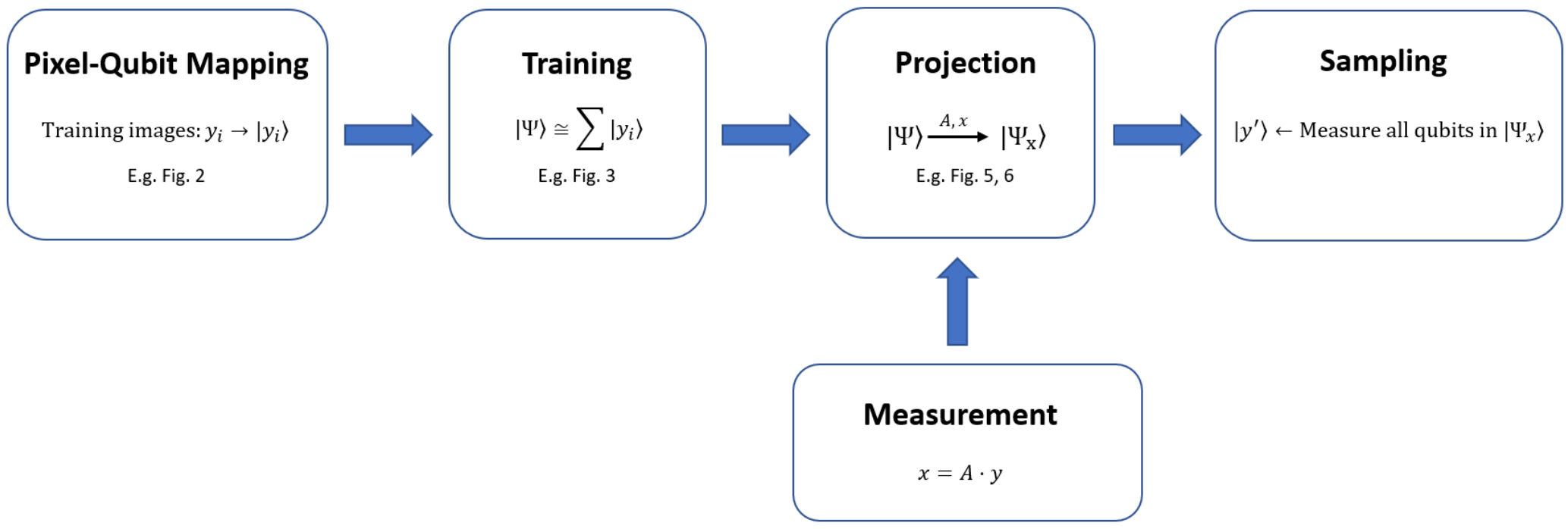

**Figure 4.** General workflow diagram of the proposed protocol where it can be seen that the Pixel–Qubit mapping (e.g., Figure 2) occurs before undertaking training (e.g., Figure 3). The projection (e.g., Figures 5 and 6) can then utilize the training and measurements which ultimately leads to the sampling.

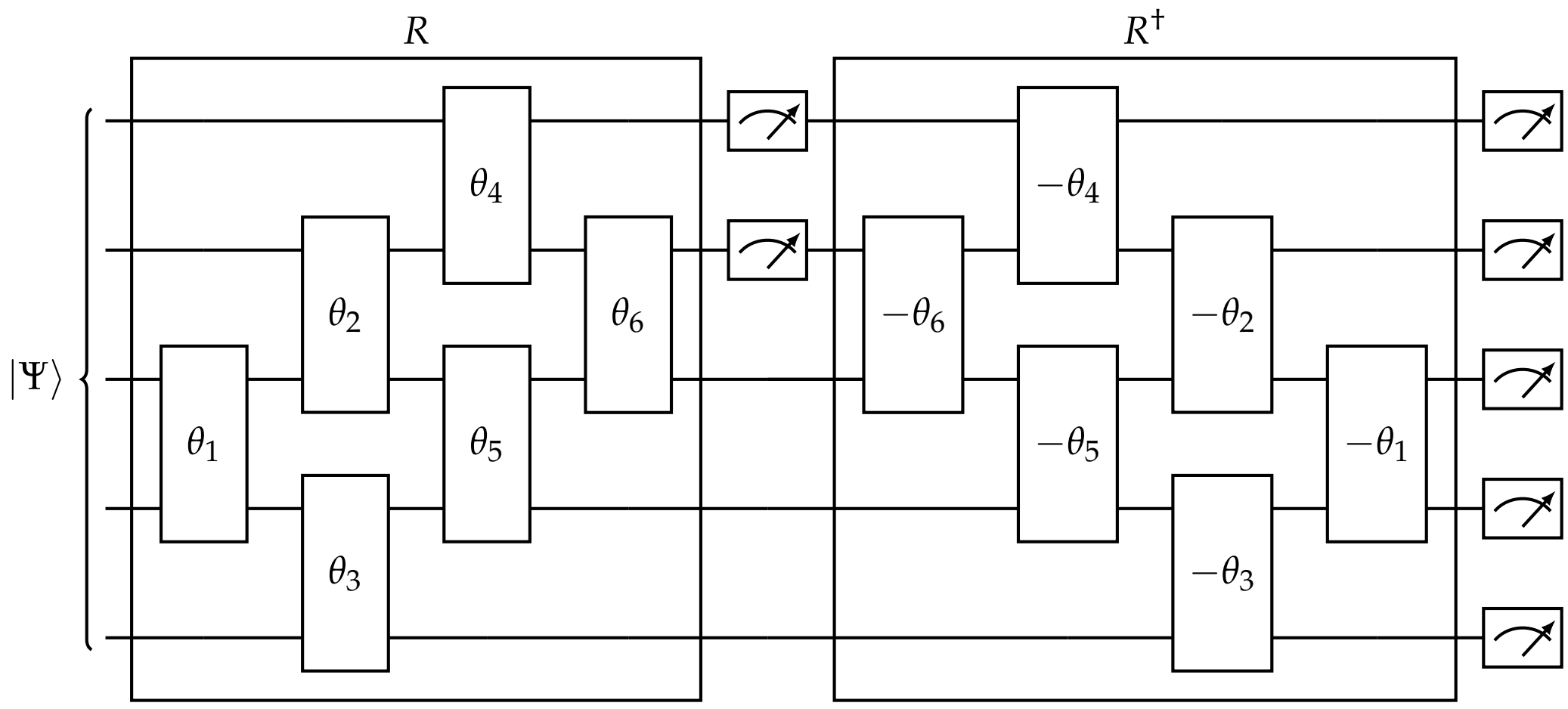

**Figure 5.** A quantum circuit to perform Quantum Compressive Sensing with the Decomposition protocol. The input state $|\Psi\rangle$ is a trained quantum state such as the one prepared by the circuit in Figure 3. Each quantum gate $\theta_i$ represents a Givens rotation, as implemented in the `cirq` Python package. Meters represent measurement, both within and at the end of the circuit.

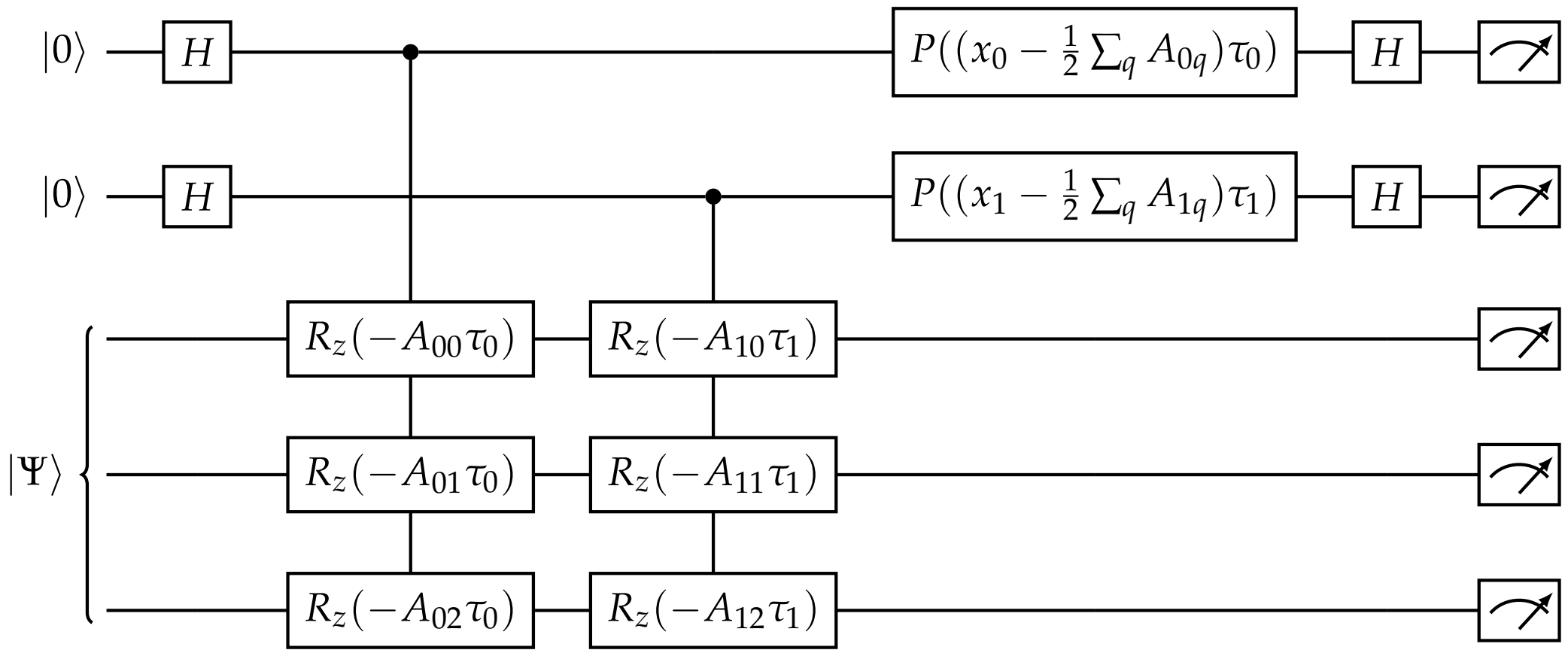

**Figure 6.** A quantum circuit to perform Quantum Compressive Sensing with the Rodeo algorithm. The input state $|\Psi\rangle$ is a trained quantum state such as the one prepared by the circuit in Figure 3. The quantum gate $H$ is the Hadamard gate. The quantum gate $R_z(\theta)$ is defined as $\exp\left(-i\frac{\theta}{2}\hat{Z}\right)$, where $\hat{Z}$ is the Pauli Z operator. The quantum gate $P(\theta)$ is the phase gate, identical to $R_z$ except by a global phase such that the $|0\rangle$ component is unchanged. Meters at the end of the circuit represent measurement.

### 5.1. Decomposition

We begin with the following decomposition for the sensing matrix $A$:

$$A = U\Delta L^\dagger \Pi R, \tag{9}$$

where $A = U\Delta V^\dagger$ is the singular-value decomposition (SVD) of $A$, and $V^\dagger = L^\dagger \Pi R$ is decomposed using a method originally developed for quantum simulation of fermionic systems [31]. Here, $L$ is another $m \times m$ unitary matrix, $\Pi$ is the $m \times n$ projection operator $\Pi \equiv \begin{bmatrix} I_m & 0 \end{bmatrix}$ representing measurement of the first three qubits, and, most critically, $R$ is an $n \times n$ unitary matrix constructed as a series of Givens rotations, which are easily adapted into a quantum circuit that can be applied to $n$ qubits. The factors $U$, $\Delta$ and $L$ are classically invertible, permitting us to define a "syndrome" s as:

$$s \equiv L\Delta^{-1}U^\dagger x = \Pi R y. \tag{10}$$

We cannot invert the singular matrix $\Pi$—this stems from the fact that there are many different $y$ which generate the same syndrome. However, the information contained in our trained Born machine $|\Psi\rangle$ is able to discriminate between them. We apply the operator $R$ as a quantum circuit which effectively rotates the computational basis of $|\Psi\rangle$ so that measuring the first three qubits yields a syndrome basis element $|s\rangle$. We repeat this procedure until $s$ and $|s\rangle$ match, at which point we rotate *back* into the original basis by applying $R^{\dagger}$. Finally, we measure all $n$ qubits to obtain a probabilistic estimate for $y$. Figure 5 shows the circuit required to execute this protocol.

Practically speaking, this protocol is suitable only for a restricted class of sensing matrices, in which $R$ is a generalized permutation matrix. To see this heuristically, consider the case where the correct image can be reconstructed deterministically *without* any classical measurements (i.e., $Y$ contains only a single binary image such that $|\Psi\rangle = |y\rangle$, a computational basis state). A well-formed projection protocol should retain this determinism, implying that the projected state $R|\Psi\rangle$ should be a factorized state $|s\rangle \otimes |\psi\rangle$, with a fixed syndrome $|s\rangle$. This is true for any $|\Psi\rangle = |y\rangle$ only when $R$ is a generalized permutation matrix. As a consequence, the only suitable sensing matrices $A$ are those which have non-zero entries in only $m$ columns (i.e., $A$ must have $n - m$ empty columns). In contrast to the original work in Ran et al. [7], which implicitly considers only sensing matrices measuring individual pixels, this is still a meaningful generalization. However, applying this protocol to fully arbitrary sensing matrices, including those used in the random sensing protocol of Duarte et al. [19], will produce useful results only to the extent that the rotation $R$ can be approximated by a generalized permutation matrix.

### 5.2. Rodeo Algorithm

Define the measurement-vector operator $\hat{N}_i$ for each row $i$ of $A$ as

$$\hat{N}_i = \frac{1}{2}\sum_q A_{iq}(\hat{I} - \hat{Z}_q), \tag{11}$$

where $\hat{I}$ is the identity operator and $\hat{Z}_q$ is the Pauli $\sigma_z$ operator on qubit $q$. When $A$ is a binary matrix, $\hat{N}$ is the standard qubit number operator acting on those pixels "measured" by $A_i$. This operator is directly connected to the measurement outcome $x_i$—in fact, we would like to project our trained Born Machine $|\Psi\rangle$ onto a quantum state $|\Psi_{x_i}\rangle$ such that the expectation value $\langle\Psi_{x_i}|(\hat{N}_i - x_i)|\Psi_{x_i}\rangle$ is 0.

We achieve this by implementing the Rodeo algorithm, a probabilistic method based on quantum phase estimation originally developed for spectrally identifying eigenvalues of a Hamiltonian operator [32]. The central idea is to apply the time-evolution operator $\hat{U} = \exp\left[-i(\hat{N}_i - x_i)\tau\right]$ to our trained Born machine $|\Psi\rangle$ for a random time $\tau$, controlled on a set of qubits in the Hadamard state. The phase accumulated by applying $U$ is "kicked back" to the control register, and can be read off directly by measuring all qubits after a *Quantum Fourier Transform*. We are interested in the case where the accumulated phase is 0, corresponding to measuring all control qubits as 0. If we successfully do so, we have (with some probability) collapsed $|\Psi\rangle$ onto $|\Psi_{x_i}\rangle$ as desired. We now repeat this procedure for all $\hat{N}_i$, and upon successfully measuring 0 on *all* control qubits, we can measure the qubits on our Born machine for a probabilistic estimate of the original signal $y$.

Our collapsed state may include any contribution accumulating a phase multiple of $2\pi$ after time $\tau$. The latter is mitigated by randomly selecting every $\tau$ from a normal distribution with variance $\sigma^2$. The contribution of phases outside of $\pm 1/\sigma$ falls off exponentially. Because $U$ in our protocol decomposes into commuting single-qubit operators, implementing its controlled variant is relatively straightforward and independent of time $\tau$ (as opposed to time-evolution of a general Hamiltonian, which may need to be decomposed into many small time steps). Therefore, $\sigma$ may be freely tuned to balance accuracy and generalizability.

Another factor affecting the success of the Rodeo algorithm is the resolution with which we can measure the phase, as determined by the number of qubits in the control register. This poses the primary limitation of the Rodeo protocol, since higher resolution requires more qubits, as well as higher qubit connectivity to accommodate controlled operations. Therefore, we adopt the highly speculative strategy of using a *single* qubit for each $\hat{N}_i$, requiring exactly $m$ auxiliary qubits (see Figure 6). The quality of any single-shot experiment is relatively low, but repeating the experiment over many trials by randomly selecting new $\tau$ for every trial has the potential to generate useful results.

### 5.3. Quantum Imaginary Time Evolution

Define a "Gaussian" operator $\hat{\Pi}_i$ as:

$$\hat{\Pi}_i = \exp\left[-\frac{(\hat{N}_i - x_i)^2}{2\sigma^2}\right], \tag{12}$$

where the measurement-vector operator $\hat{N}_i$ is defined as in Equation (11). Because $\hat{\Pi}_i$ is diagonal, its action on a computational basis state $|z\rangle$ can be written as

$$\hat{\Pi}_i|z\rangle = \exp\left[-\frac{(\langle z|\hat{N}_i|z\rangle - x_i)^2}{2\sigma^2}\right]|z\rangle. \tag{13}$$

Therefore, applying $\hat{\Pi}_i$ on our Born machine $\Psi$ exponentially suppresses any basis state whose expectation value $\langle \hat{N}_i \rangle$ differs from $x_i$ considerably. Repeating this procedure for each $\hat{N}_i$, we measure all qubits, and obtain a probabilistic estimate of the basis vector $y$ most consistent with all measurement outcomes $x_i$. Because $\langle z|\hat{N}_i|z\rangle$ can only take discrete values, we require a finite variance $\sigma^2$ to accommodate all measurement outcomes; $\sigma = 0.5$ is a sensible choice, although it may be tuned to balance accuracy and generalizability.

This approach is especially attractive because there is no syndrome measurement—success is "guaranteed". However, the Gaussian operator $\Pi_i$ is neither unitary nor even idempotent—there is no way to implement it on an arbitrary quantum state. Instead, we may use the methods of Quantum Imaginary Time Evolution (QITE) to implement a unitary operator which *effectively* realizes the same operation (after re-normalization), for our *particular* input state. There are two major technical difficulties in the formulation of QITE used by Motta et al. [33]. First, quantum state tomography over $|\Psi\rangle$ is required to construct an effective Hamiltonian $\hat{H}$, and second, the time-evolution $e^{-i\hat{H}\delta\tau}$ must then be implemented over many small time-steps via Suzuki–Trotter decomposition. Because our operator $\Pi_i$ is diagonal, it may have a straightforward implementation. However, because it entangles all qubits, this implementation still scales exponentially. In this work, we steer clear of these difficulties via implementing $\Pi_i$ only by direct classical simulation (i.e., calculating $\langle z|\Pi_i|z\rangle$ for all $|z\rangle$). Nevertheless, QITE is an active area of research and we judge this method to be the most likely to yield favorable scaling in the long-term.

### 5.4. Entangled Born Machine

A bottleneck in the application of both the Decomposition and Rodeo projection protocols is the number of trials required to obtain the appropriate syndrome measurement. For any given $s$ and $y$, the probability $p$ to measure $|s\rangle$ varies. Any one instance of our protocol requires a number of trials scaling with $p^{-1}$. The expected value of the number of trials is a weighted average over all possible $|s\rangle$:

$$\langle N_{\text{trials}} \rangle = \sum_{|s\rangle} p p^{-1} = 2^m. \tag{14}$$

Thus, the number of trials is exponential with the number of measurements—an extremely unfavorable scaling. Worse yet, rarer signals that coincide with the most interesting features would require even more trials.

In light of the above, one important area of further research is to identify deterministic projection protocols. One possible line of inquiry is to consider multiple, entangled copies of $\Psi$. For example, assume $|\Psi\rangle$ is in the following state:

$$|\Psi\rangle = a\,|0\rangle|\Psi_x\rangle + b|1\rangle|\Psi_1\rangle. \tag{15}$$

Further, assume that our target syndrome is $|s\rangle = |0\rangle$, and we aim to isolate $|\Psi_x\rangle$. Given this state, we can only hope that measuring the first qubit yields 0, as in the first two projection protocols described above. However, given the entangled state

$$|\Phi\rangle = \frac{a+b}{\sqrt{2}}|0\rangle|\Psi_x\rangle|\Psi_1\rangle + \frac{a-b}{\sqrt{2}}|1\rangle|\Psi_1\rangle|\Psi_x\rangle \tag{16}$$

one can measure the first qubit: if it is 0, $|\Psi_x\rangle$ is prepared in the first copy; if it is 1, we take the second copy instead. This method requires just one trial, and guarantees success. However, it instead requires an exponential number of *copies*, and it is not clear how to prepare the entangled state $|\Phi\rangle$, given a protocol to prepare the single state $|\Psi\rangle$. We leave further investigation of this approach to future work.

## 6. Demonstration

We are currently in the Noisy Intermediate-Scale Quantum (NISQ) era, which means that the quantum computers available today have relatively few qubits, and require complex error mitigation protocols far beyond the scope of this paper. Therefore, we opt to demonstrate our algorithm via a classical simulation of very small, six-pixel signals.

### *6.1. Description of Model*

We consider a qualitative model of forest foliage that may be measured by overhead LIDAR imaging missions, such as the Concurrent Artificially-intelligent Spectrometry and Adaptive Lidar System (CASALS) satellite currently under development at NASA. Our model consists of two categories: a relatively young forest (Figure 7a), in which the canopy is still rising, and a mature forest (Figure 7b), in which the canopy has attained its maximum height and underbrush is measurably dense. We emphasize that this model is purely qualitative, and contrived solely for the sake of demonstrating our Quantum Compressive Sensing protocol on a tractably-sized system. In particular, the resolution attainable in real-world LIDAR data is multiple orders of magnitude greater than in the model presented here.

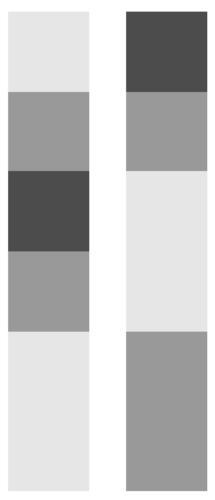

(**a**) (**b**)

**Figure 7.** Prototypical images in a qualitative model for young (**a**) and mature (**b**) foliage. Training and validation signals are generated as random perturbations on one of these two images.

The model is sampled by randomly selecting the class (young or mature), then perturbing each pixel in the prototype images of Figure 7 with Gaussian noise representing variation in the distribution of foliage. We construct three separate training sets sampling sixteen images each, shown in Figure 8.

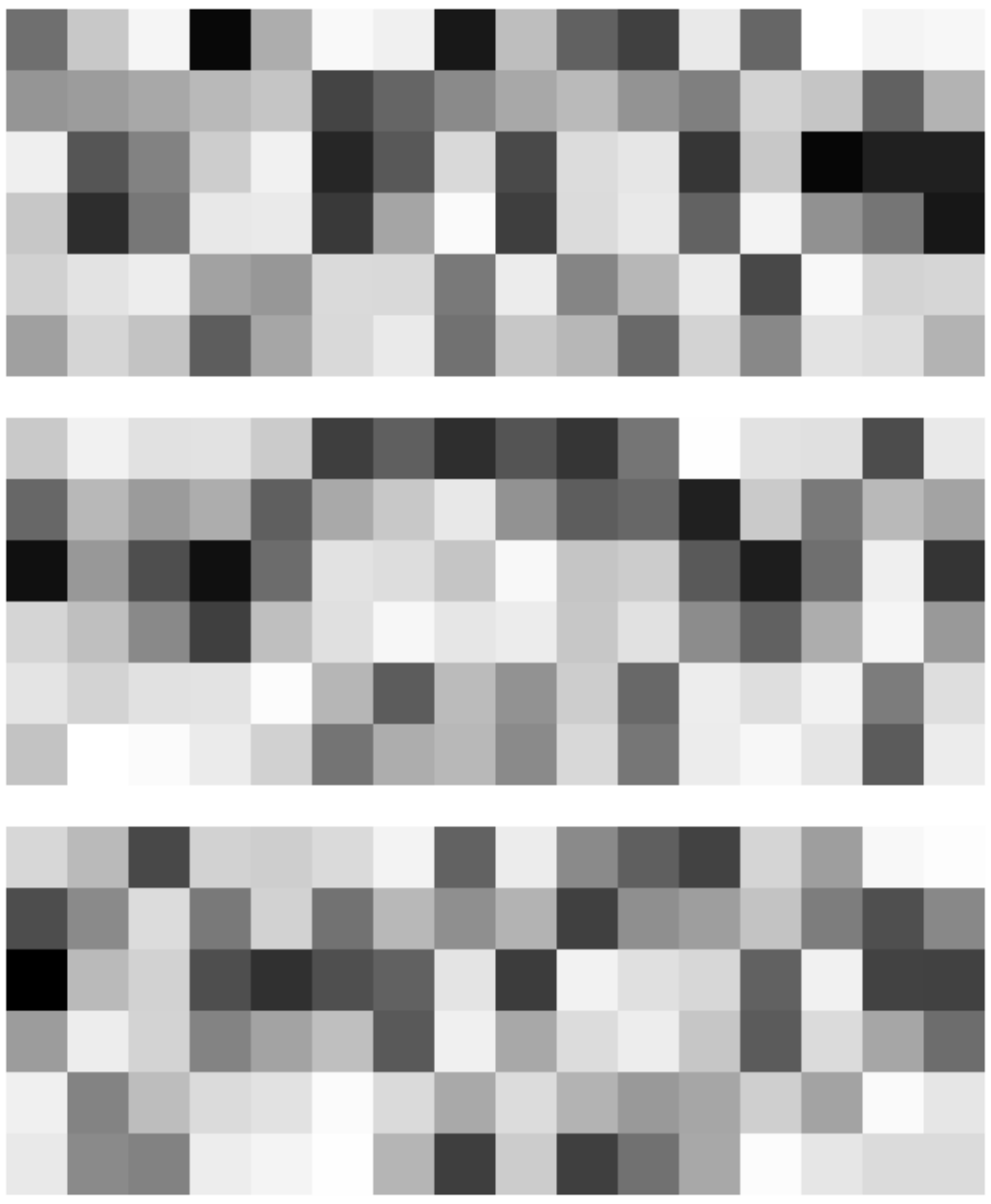

**Figure 8.** Three training sets generated from the prototype signals shown in Figure 7.

Using the "quantum average" training protocol discussed in Section 4, each training set is mapped deterministically onto a pure quantum state $|\Psi\rangle$, depending only on the parameter $p$ used for the qubit mapping (Equation (2)). We consider an "optimal" $p$ for each training set which *maximizes* the information entropy $S$ of the resulting probability distribution $P(z)$. Figure 9 shows the dependence of $S$ on the parameter $p$ for each training set. We select $p = 0.5$ as a "default" choice and plot the probabilities $P(z) \equiv |\langle z|\Psi\rangle|^2$ of measuring each basis vector $|z\rangle$ for each training set in Figure 10a, while Figure 10b plots the new probability distributions when the $|\Psi\rangle$ are formed using the optimal $p$. The gray bars mark the probabilities of the "ideal" training set constructed precisely from the prototype images in Figure 7. In all three cases, the "quantum average" successfully produces results which are qualitatively similar to one another and to the ideal, although we can see in Figure 9 that averaging over random perturbations from the ideal has the effect of smoothing out potentially interesting features in the overall signal space.

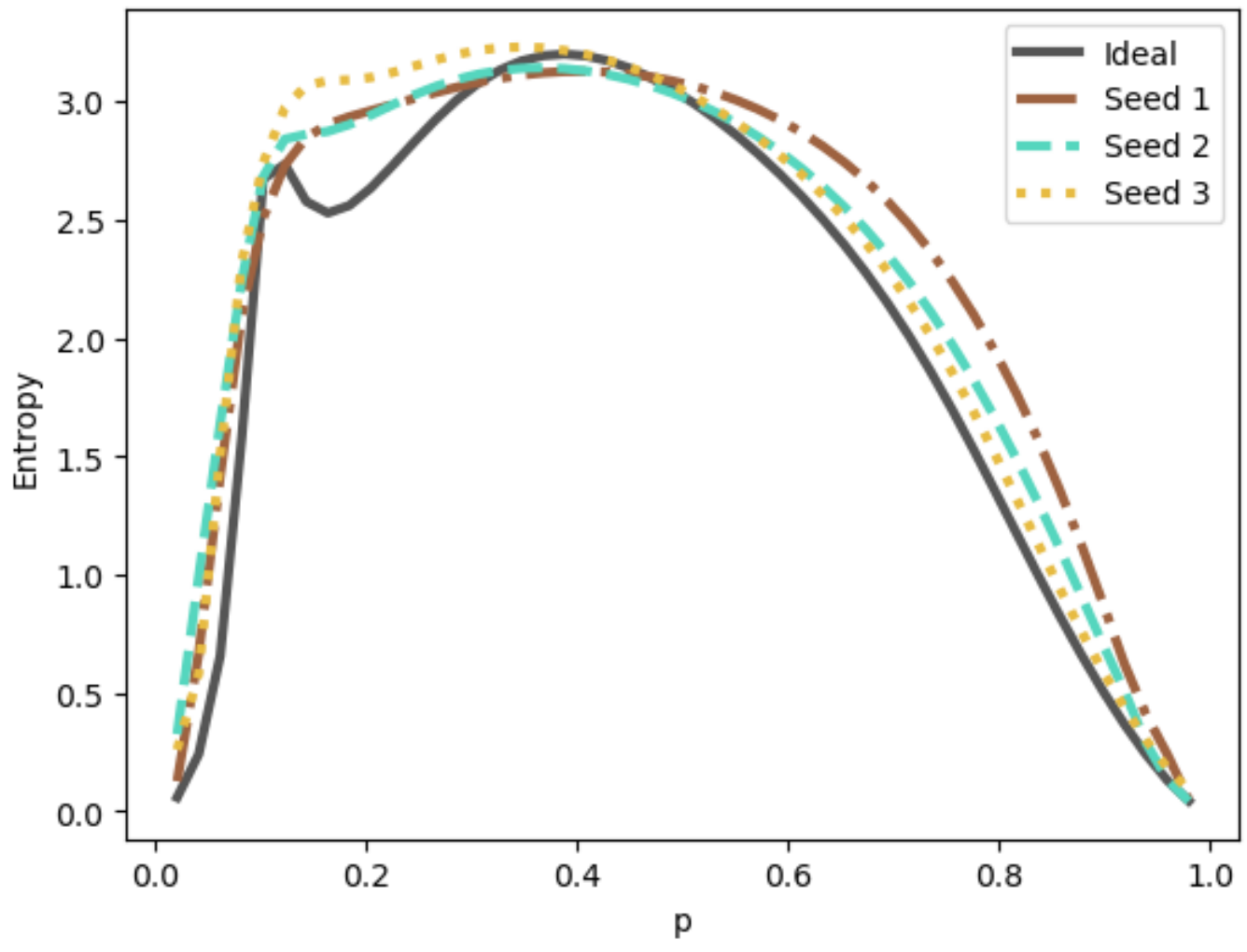

**Figure 9.** Information entropy of the training set as a function of the "midpoint" $p$ used in Equation (2). The ideal curve considers a set of images which are split evenly between the two prototype signals.

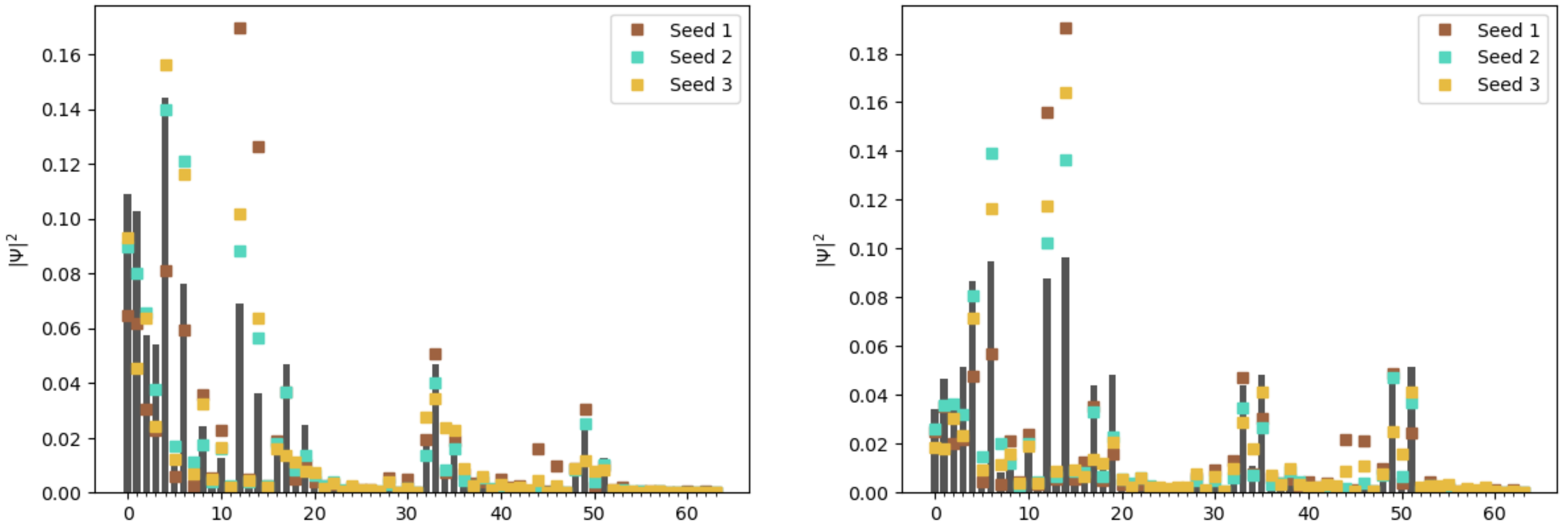

(**a**) Default ($p = 0.5$) (**b**) Maximized Entropy

**Figure 10.** Square amplitude of each basis vector for an ideal state (bars) and each trained state (squares). (**a**) uses a "midpoint" value of $p = 0.5$ while (**b**) optimizes $p$ for each state to maximize the information entropy, as visualized in Figure 9.

Throughout this section, we will assess the accuracy of our Quantum Compressive Sensing protocol using a measure we refer to as the "Relative Log Likelihood" (RLL), derived in Appendix A. The RLL is a function of both the probability $P(z)$ that a binary image $z$ is generated by a given signal $y$, and the information entropy $S$ intrinsic to that signal.

To help visualize it, we plot the RLL of all possible $z$ with respect to the signals $y$ corresponding to the prototype images in Figure 7, using both the default $p$ (Figure 11a) and the optimal $p$ (Figure 11b). In general, a higher RLL indicates greater success, but we consider any RLL relatively close to zero (even negative values) to be "good". For example, the binary image corresponding to $z = 3$ is $|000011\rangle$, evidently corresponding to a tall, thick canopy (reading the left-most bit as closest to the ground), consistent with a mature forest. Its RLL value (using optimal $p$) is 0.51. The binary image $|000010\rangle$ corresponding to $z = 2$ is also consistent with a mature forest and has an RLL of $-1.24$. That said, because the canopy is less thick, not quite as tall, and there is no immediate sign of underbrush, it is perhaps slightly more consistent with a young forest, as captured by the ever so slightly higher RLL of $-1.19$.

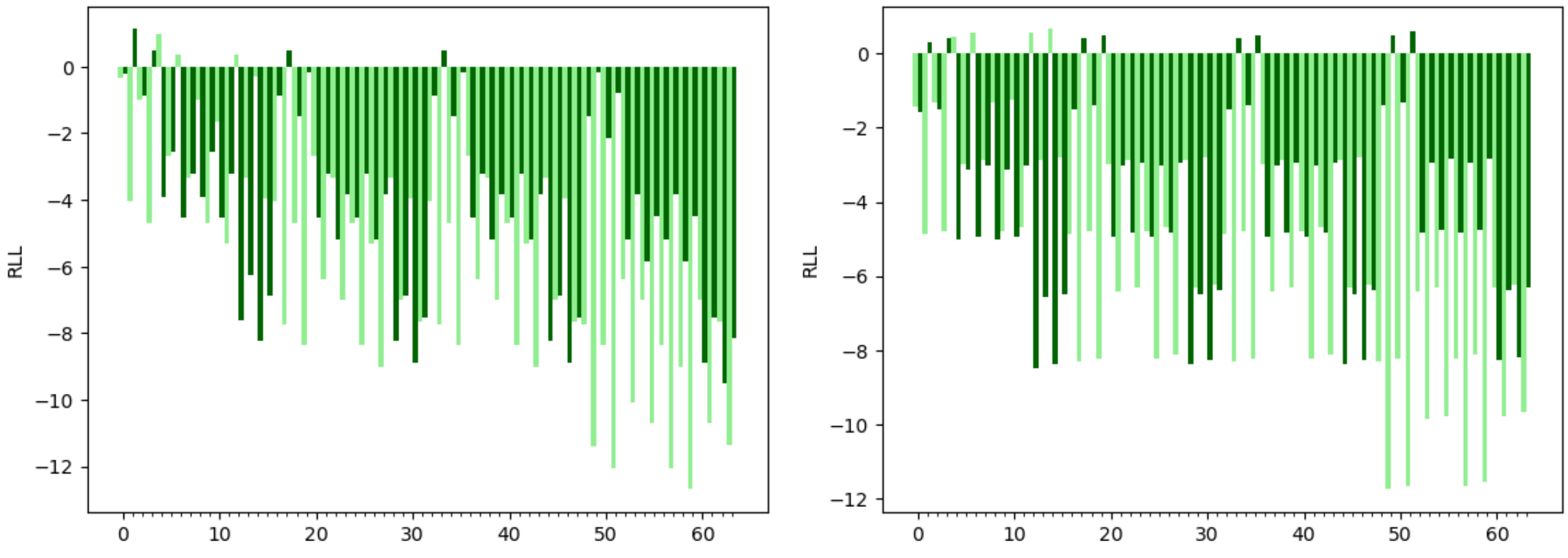

(**a**) Default ($p = 0.5$) (**b**) Maximized Entropy

**Figure 11.** Relative log likelihood of each basis vector for an ideal state. (**a**) uses a "midpoint" value of $p = 0.5$ while (**b**) optimizes $p$ for each state to maximize the information entropy, as visualized in Figure 9.

### 6.2. Quantum Compressive Sensing

We now present the results of our Quantum Compressive Sensing protocol, applied to our qualitative model. We begin each trial by preparing the Born machine $|\Psi\rangle$, using one of the training sets constructed above, or the ideal based on the prototype images in Figure 7. We then generate a random signal $y$ from the forest model, and a random $m \times n$ sensing matrix $A$. We calculate $x = Ay$, and then implement one of the protocols presented in Section 5. The protocol ends by measuring all qubits, yielding a binary image $z$. Each trial is repeated 1024 times for the same signal $y$ and sensing matrix $A$ (but generally different $z$) to obtain the median RLL relating $z$ and the original signal $y$. Finally, this process is repeated for 32 different $y$ and $A$ to obtain a distribution of median RLL (we use the same $y$ and $A$ for all experiments). Throughout this section, we plot the inter-quartile ranges of these distributions as a function of the number of measurements $m$. If our Quantum Compressive Sensing is successful, we should observe an increase in median RLL as $m$ increases.

Figure 12a shows our results when directly applying the Gaussian operator in Equation (12) with $\sigma = 0.5$ to each of the training sets, constructed using the default $p$. Each sensing matrix $A$ is a binary matrix with equal probability between 0 and 1, consistent with the micro-mirror LIDAR design presented in Duarte et al. [19]. We remind the reader that we are applying the Gaussian operator for this demonstration, but this can in principle be done in a quantum way with Quantum Imaginary Time Evolution. Hereafter, we label this projection protocol as QITE. This protocol will clearly be successful: At $m = 0$, we are essentially using our Born machine for generative sampling of our model, generating a binary image $z$ which is consistent with either a young or mature forest, but with no preference based on the original signal $y$. Those $z$ which happen to match the class (that is, young or mature) of $y$ are likely to have acceptable RLL, but we are just as likely to generate $z$ with very poor RLL. As $m$ increases, however, the projection step "teaches" the Born machine more and more about the original signal $y$, enabling us to properly distinguish between the two classes and even select an optimal image among them.

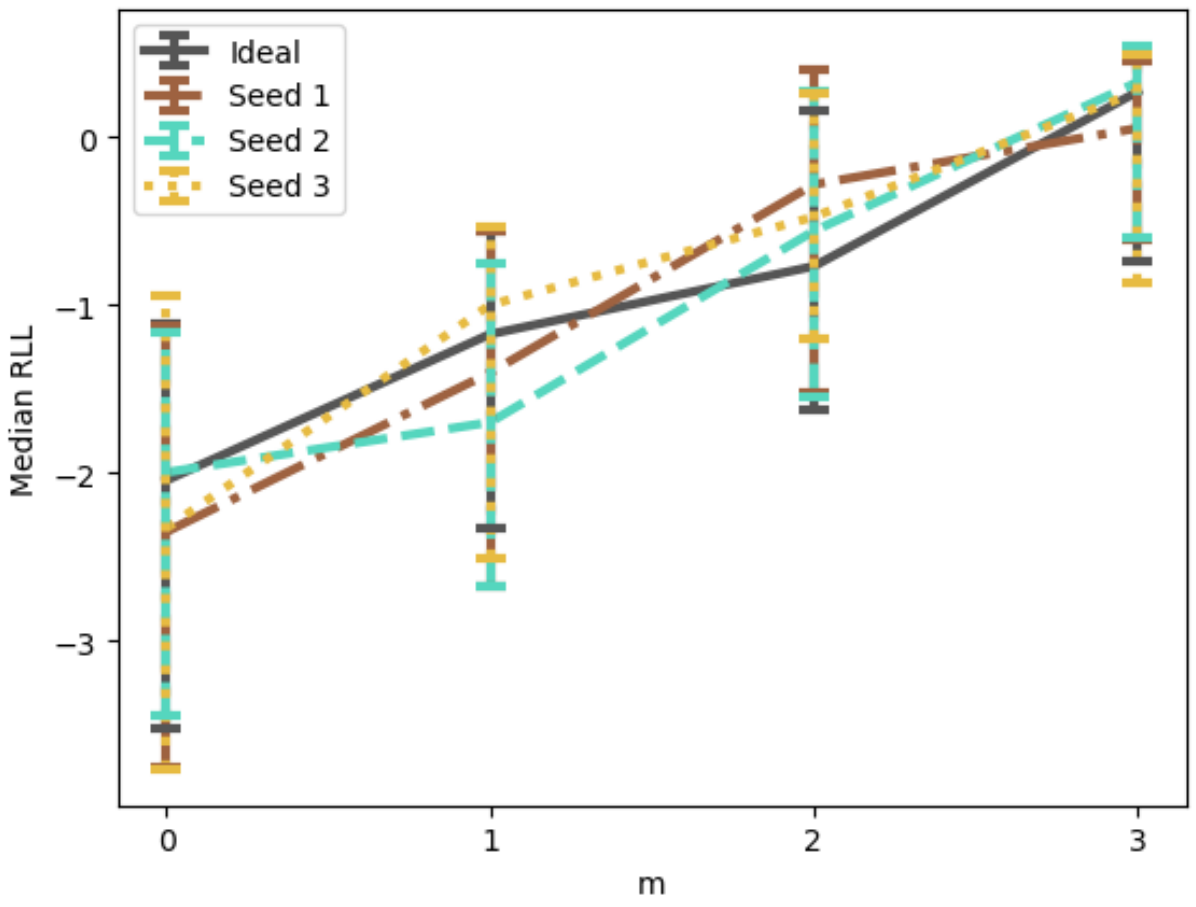

(**a**) QITE protocol, contrasted over training sets.

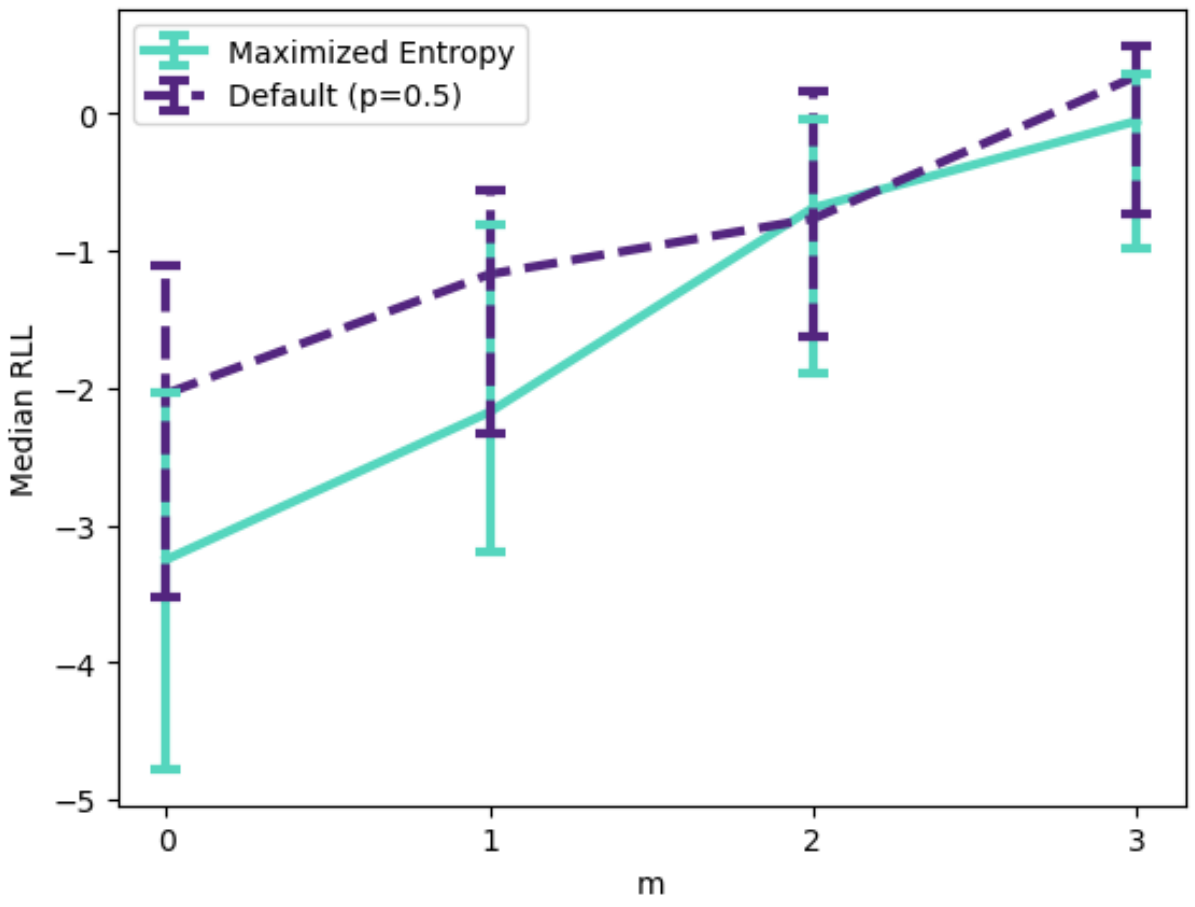

(**b**) QITE protocol, contrasted over midpoints.

**Figure 12.** Median relative log likelihood as a function of measurements, after applying the QITE protocol over 1024 trials. (**a**) contrasts performance for each training set, while (**b**) contrasts different choices of the "midpoint" $p$. At $m = 0$ in (**a**), our Born machine is being utilized for generative sampling of our model. As $m$ increases, the projection step "teaches" the Born machine about the original signal $y$, allowing for a distinction between the two classes and optimal image selection. In (**b**), we see that the default choice $p = 0.5$ produces better results and we thus utilize $p = 0.5$ for all our experiments.

In Figure 12b, we investigate how our choice of $p$ in constructing the trained Born machine $|\Psi\rangle$ influences results. We focus on just the ideal training set and continue to use the QITE protocol with $\sigma = 0.5$. In our simulations, the default choice $p = 0.5$ turns out to produce better results, at least for this choice of experimental parameters. Therefore, for the remainder of this section, we set $p = 0.5$ for all experiments. This opens the question whether an alternative choice of $p$ may improve results further. For example, while we justify the choice of a single $p$ for the entire image by thinking of it as a normalization on the measuring apparatus, we might equally well optimize $p$ for each individual pixel. We leave this line of inquiry as a direction for future research.

The QITE projection protocol is unique in that it is deterministic, in the sense that the state of the Born machine $|\Psi\rangle$ is fixed prior to sampling. In the Rodeo and Decomposition protocols, the state of $|\Psi\rangle$ is contingent upon measurements made prior to sampling. Indeed, the state of $|\Psi\rangle$ is "correct" only when the correct syndrome is measured ($|\mathbf{0}\rangle$ for Rodeo, and $s$ from Equation (10) for Decomposition). If this syndrome is *not* measured, the entire trial must be discarded. In Figure 13a, we find the number of failures out of the 1024 trials for each signal $y$, and we plot the inter-quartile ranges of the distribution of failures counts over all 32 signals. We anticipate the chance of failure to increase with $m$, since larger $m$ have more complicated syndromes, which is also observed here.

We also observe in Figure 13a that the chance of failure increases with the size of $\sigma$ used in the Rodeo protocol. Smaller $\sigma$ mean the random times $\tau$ are relatively small, inducing relatively little phase shift on the Born machine $|\Psi\rangle$. As such, phase kickback is more likely to collapse to the smallest value 0. This is beneficial for obtaining the correct syndrome, but it does not "cast the net" wide enough, reducing the likelihood to project onto the desired subspace. On the other hand, increasing $\tau$ to arbitrarily large values may "cast the net" so wide that our limited number of trials (32) is insufficient to adequately sample the whole space. Thus, for any fixed number of trials, we can expect an optimal choice of $\sigma$ which maximizes the increase in median RLL, as observed in Figure 13b.

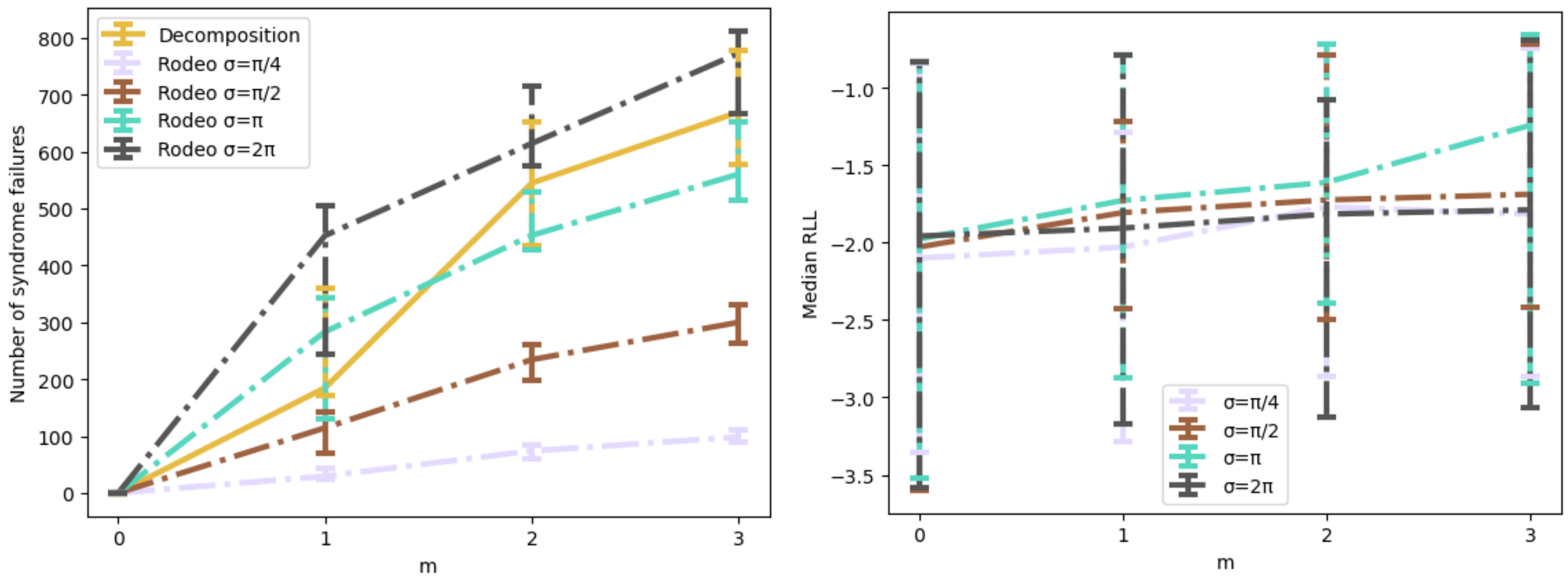

(**a**) Syndrome failures (out of 1024 attempts)

(**b**) Rodeo protocol, contrasted over $\sigma$

**Figure 13.** A comparison of the Rodeo protocol with different choices of $\sigma$. (**a**) shows the number of syndrome failures out of 1024 attempts, and (**b**) shows the median relative log likelihood for successful attempts. (**a**) also includes failures for the Decomposition protocol, for comparison. We clearly observe in (**a**) that the chance of failure increases with the size of $\sigma$ used in the Rodeo protocol and can see in (**b**) that we can expect an optimal choice of $\sigma$ which maximizes the increase in median RLL.

As our last demonstration, we compare the performance of each projection protocol (Decomposition, QITE with $\sigma = 0.5$, Rodeo with $\sigma = \pi$) under different classes of sensing matrix. Figure 14a considers the same binary sensing matrices we have used thus far, while Figure 14b considers sparse binary sensing matrices where the probability of each 1 is only 20%. We immediately observe that the Decomposition protocol *lowers* the accuracy of reconstruction when measurements are introduced. This is anticipated as the considered class of sensing matrices does not satisfy the requirement that $A$ is non-zero on only $m$ columns. That said, the sparse matrices (Figure 14b) better approximate this condition, and we see that the Decomposition protocol actually performs reasonably well for larger $m$. Meanwhile, the Rodeo algorithm underperforms for both classes of matrices-while the positive trend is evident, and it is more consistent than the Decomposition protocol for small $m$, it has a significantly smaller slope of enhancement than its competitors. We see this trend continue in its extreme in Figure 14c, which considers the simple case where each row of $A$ measures a single pixel. The Rodeo algorithm is a stochastic algorithm heavily reliant on sampling over a representative distribution of phases, and it may perform better when the "one-shot" sampling approach used in this paper is relaxed.

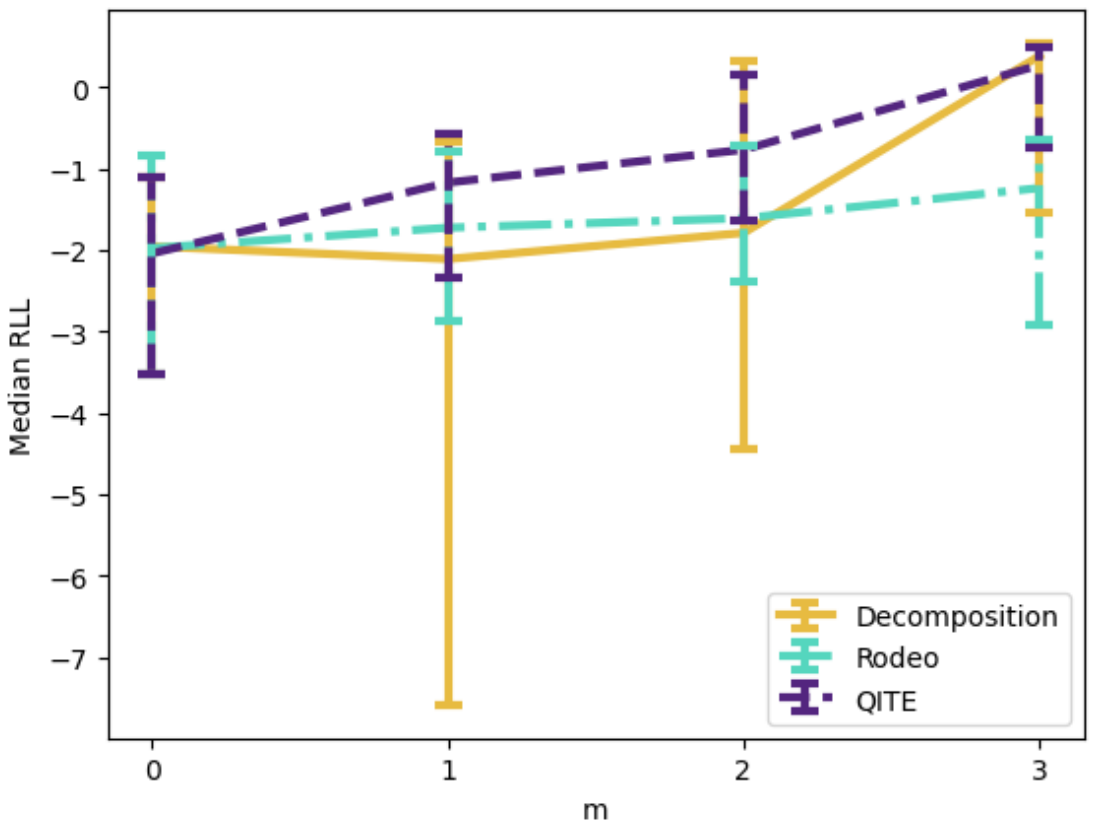

(**a**) Binary $A$: $A_{ij}$ is 0 or 1 with equal probability.

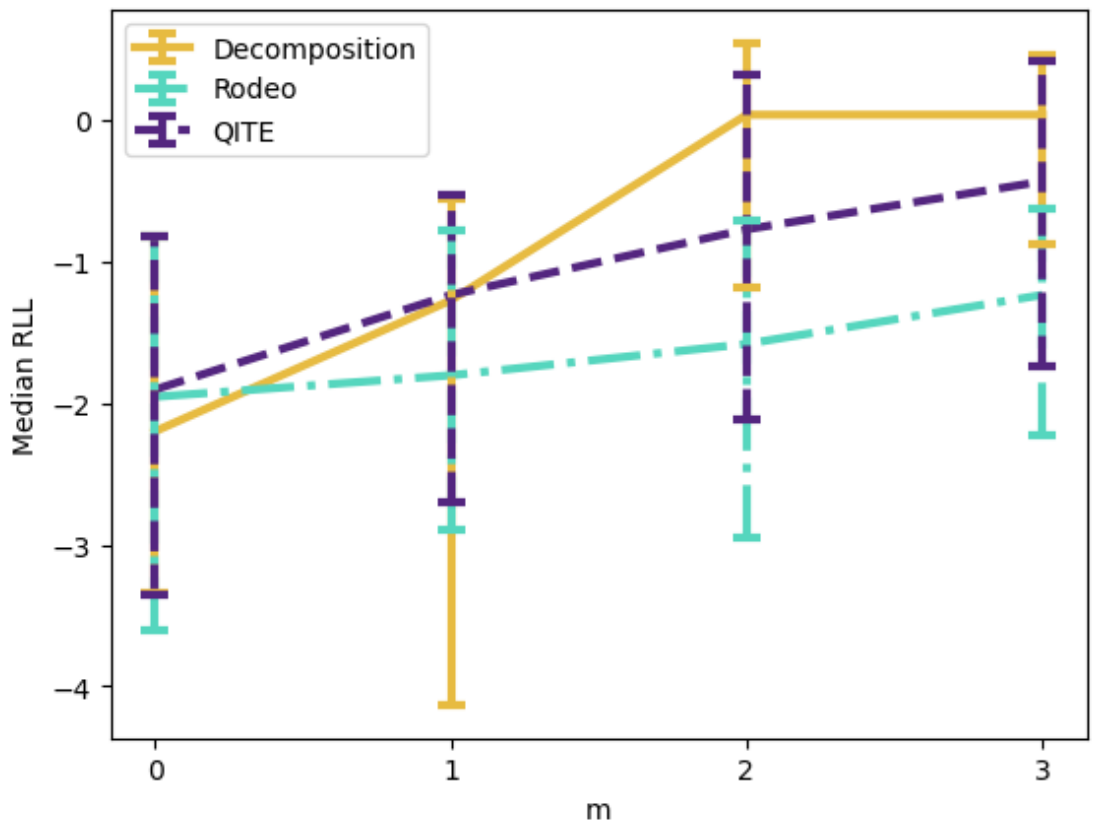

(**b**) Binary $A$: $A_{ij}$ is 1 with 20% probability.

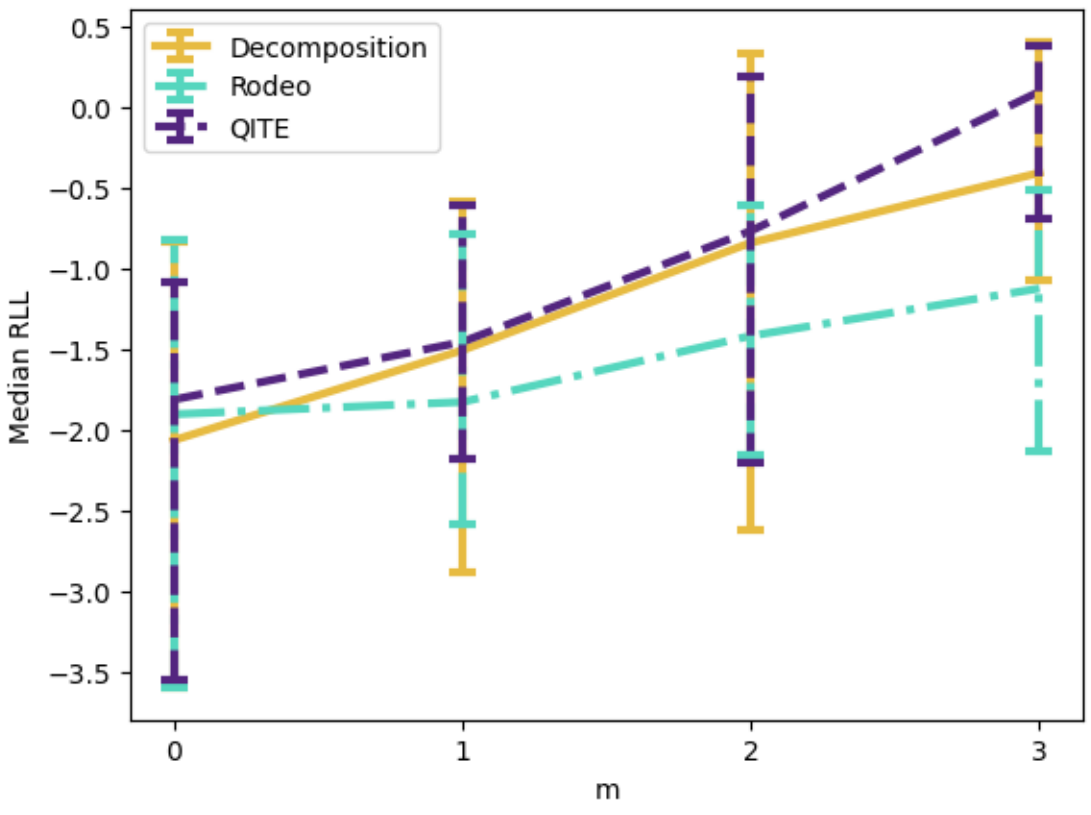

(**c**) Binary $A$: each row of $A$ has only a single 1.

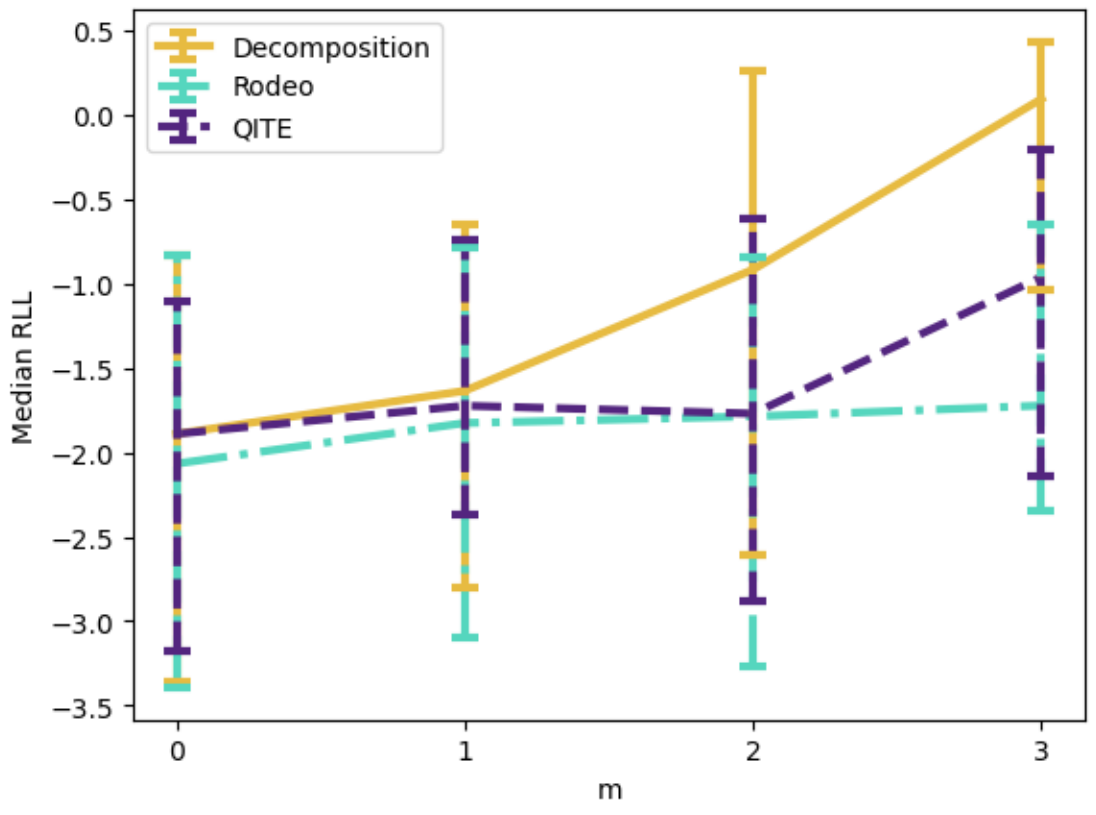

(**d**) Uniform $A$: only $m$ columns of $A$ are nonzero.

**Figure 14.** Comparison of each Quantum Compressive Sensing protocol, for different classes of sensing matrix. (**a**), (**b**), (**c**) utilize a binary $A$ with $A_{ij}$ being 0 or 1 with equal probability, a binary $A$ with $A_{ij}$ being 1 with 20% probability, and a binary $A$ with each row of $A$ having only a single 1 respectively. (**d**) utilizes a uniform $A$ where only $m$ columns of $A$ are nonzero. QITE uses $\sigma = \frac{1}{2}$ and Rodeo uses $\sigma = \pi$. In (**a**,**b**) we see (as anticipated) that the Decomposition protocol lowers the accuracy of reconstruction when measurements are introduced, since these classes of sensing matrix do not satisfy the condition that $A$ is non-zero on only $m$ columns. However, (**b**) and especially (**c**) demonstrate that sparse matrices can approximate this condition, increasing accuracy of the Decomposition protocol for larger $m$. In (**d**), the performance of QITE and Rodeo fall off due to the restrictions of the sensing matrix, but the Decomposition performance is impressive.

Finally, Figure 14d considers non-binary matrices in which $m$ columns are sampled uniformly between 0 and 1, and the remaining $n - m$ columns are left at zero. These sensing matrices are inherently flawed, in that they are guaranteed to only ever measure a restricted portion of the signal $y$. Any distinguishing characteristics found in the unmeasured portion of the signal cannot be used to project the Born machine $|\Psi\rangle$. Indeed, we see in this case that the QITE protocol, up to this point the gold standard, falls off significantly, and the rodeo algorithm fails all but completely. That said, this class of sensing matrices is the most general on which we can expect the Decomposition protocol to function as intended, and indeed we observe that it performs extremely well. We conclude that Quantum Compressive Sensing, implemented with a projection protocol suited to the particular sensing matrix in question, is an effective means of introducing incomplete information associated with a signal of interest.

## 7. Conclusions

In this paper, we have presented a protocol for performing "quantum" compressive sensing, and suggested several ways in which each step of the protocol may be implemented on a quantum computer. The protocol begins by pre-preparing a set of qubits into a quantum state $|\Psi\rangle$ which captures the statistics of all potential signals. The quantum state $|\Psi\rangle$ is then evolved with a series of quantum operations which cause $|\Psi\rangle$ to collapse into a lower-dimensional sub-space consistent with the compressed measurement made on the signal. Finally, the qubits are measured, producing a probabilistic estimate of the original image. Our formalism generalizes to all linear sensing matrices, and we have presented a preliminary demonstration of our method on a very small, classically-simulable model.

The algorithms discussed in this paper all depend on a one-qubit-per-pixel approach, relegating the application of these results to meaningfully-sized images to far in the future, when quantum computers with hundreds or thousands of qubits are available. We have also not contended with the formidable technical challenge of mitigating noise inherent to all quantum computing systems available today. Developing alternative implementations with a more compact pixel–qubit mapping would be an excellent direction for further research. For example, in the case where the signal basis is indeed sparse, as in the classical compressive sensing paradigm, one may assign $\log(n)$ qubits to each of $k$ "on" bits, indexing at which pixel its particular "on" bit is found (for a total of $k \log n$ qubits). The quantum chemistry literature, from which we have borrowed the majority of our techniques, names this the "first-quantization" approach, and developing efficient algorithms for navigating this qubit mapping is an area of active research.

In the long term, when large, error-corrected quantum computers are readily available, the compressive sensing methods in this paper are expected to be useful for processing high-dimensional, high-resolution signals. Thus, they have wide-ranging applications to earth science, astrophysics, communications, and beyond. Moreover, the classical constraint requiring a basis in which the signal is known to be sparse is no longer necessary. Instead, we use a data-driven approach, using a representative sample of signals in the basis most convenient for representing the sensing matrix. This promises to be a very powerful perspective in the quantum implementation of the compressive sensing literature.

**Author Contributions:** Conceptualization, K.M.S., N.N., H.S., H.C.S. and M.S.; methodology, K.M.S. and N.N.; software, K.M.S. and N.N.; validation, K.M.S., N.N., H.S., H.C.S. and M.S.; formal analysis, K.M.S., N.N. and H.S.; investigation, K.M.S. and N.N.; resources, H.S., H.C.S. and M.S.; data curation, K.M.S.; writing—original draft preparation, K.M.S.; writing—review and editing, N.N., H.S., H.C.S. and M.S.; visualization, K.M.S. and N.N; supervision, H.S., H.C.S. and M.S.; project administration, H.S., H.C.S. and M.S.; funding acquisition, H.S., H.C.S. and M.S. All authors have read and agreed to the published version of the manuscript.

**Funding:** Funding for this research was provided by the NASA Goddard Space Flight Center Advanced Communications Capabilities for Exploration and Science Systems (ACCESS) Project Office. This research was also partially funded by the National Reconnaissance Office and the Universities Space Research Association.

**Institutional Review Board Statement:** Not applicable.

**Informed Consent Statement:** Not applicable.

**Data Availability Statement:** Not applicable.

**Acknowledgments:** We thank Asmita Korde-Patel, James MacKinnon, and David Harding for useful discussions.

**Conflicts of Interest:** The authors declare no conflict of interest. The funders had no role in the design of the study; in the collection, analyses, or interpretation of data; in the writing of the manuscript; or in the decision to publish the results.

## Abbreviations

The following abbreviations are used in this manuscript:

| | |
|---|---|
| MPS | Matrix Product State |
| TNCS | Tensor Network Compressed Sensing |
| LIDAR | Light Detection and Ranging |
| CASALS | Concurrent Artificially-Intelligent Spectrometry |
| NLL | Negative Log Likelihood |
| SVD | Singular-Value Decomposition |
| QITE | Quantum Imaginary Time Evolution |
| NISQ | Noisy Intermediate-Scale Quantum |
| RLL | Relative Log Likelihood |

## Appendix A. Relative Log Likelihood

In this section, we develop a bias-free measure for rating how well an image *estimate* $|z\rangle$ matches the true signal $y$, suitable for contrasting performance across multiple trials and implementations. With a "one-shot" sampling approach, the output $|z\rangle$ of our projection protocol is always a computational basis state:

$$|z\rangle = \bigotimes_{q=1}^{n} |z_i\rangle, |z_i\rangle \in \{|0\rangle, |1\rangle\}. \tag{A1}$$

Meanwhile, our pixel–qubit mapping and training protocol allow for gray-scale signals $y = [y_0 ... y_n]$, $y_i \in [0, 1]$. The best possible binary image is straightforward to derive from the gray-scale signal, given the mapping in Equation (2): select $|z_i\rangle = |0\rangle$ if $y_i < p$ and $|z_i\rangle = |1\rangle$ otherwise. However, even when the correct signal $y$ is successfully discriminated by the compressive sensing protocol, the probability that the final readout step produces the optimal image is determined by the precise values of $y_i$, and will in general be very low as $n$ increases. We should consider $|z\rangle$ "correct" as long as it remains qualitatively similar to $y$. Our goal in this section is to develop a *quantitative* statistic for the similarity between $|z\rangle$ and $y$.

As a first attempt, consider the fidelity $F$:

$$F(y, z) \equiv |\langle z|y\rangle|^2, \tag{A2}$$

where $|y\rangle$ is the quantum state mapped by Equation (2). This is an intuitive choice and a common metric for determining distance between quantum states. However, it is not *fair*, in the sense that different $y_i$ have drastically different distributions of $F$. To quickly appreciate this, consider the two-dimensional signals $y_a = \begin{bmatrix} 1 & 0 \end{bmatrix}$ and $y_b = \begin{bmatrix} \frac{2}{3} & \frac{1}{3} \end{bmatrix}$, and set the midpoint $p = 0.5$. For both signals, the best possible binary image is $|z\rangle = |10\rangle$. Finding the fidelity of $|z\rangle$ with respect to each state, we see $F_a = 1$ while $F_b = \frac{9}{16}$. This shows that the best possible fidelity for $y_b$ is barely half that of $y_a$. Evidently, fidelity alone is not suitable for contrasting performance across multiple signals.

The measure we propose is the following:

$$RLL(y,z) = \ln F(y,z) + S(y), \tag{A3}$$

christened RLL for "Relative Log Likelihood", after its closely related cousin the Negative Log Likelihood (see Equation (4)). The first term $\ln F$ is simply the fidelity from the previous paragraph, scaled logarithmically. The second term $S$ normalizes the log fidelity by subtracting off the *average* log fidelity over all possible choices of $|z\rangle$, weighted by the probability of obtaining $|z\rangle$ upon measuring $|y\rangle$. Naturally, this probability is precisely the fidelity $F$, and $S$ turns out to be the information entropy of the distribution over all $F$, i.e.,

$$S(y) = -\sum_z F(y,z) \ln F(y,z). \tag{A4}$$

Because entropy is an extrinsic property, we know that $S(y)$ can be calculated efficiently pixel by pixel via $S(y) = \sum_{i=1}^n S(y_i)$, where the information entropy for a single pixel is a simple sum over $z_i = \{0,1\}$.